\documentclass[aps,prl,groupedaddress,twocolumn]{revtex4}%
\usepackage{amsfonts}
\usepackage{amsmath}
\usepackage{graphicx}
\usepackage{amssymb}%
\providecommand{\U}[1]{\protect\rule{.1in}{.1in}}
\begin{document}
\title{A common model for superluminal propagation in absorbing atomic media and lossy metastructures}
\author{Pedro Chamorro-Posada}
\email{pedro.chamorro@tel.uva.es}
\affiliation{Dept. Teor\'{\i}a de la Se\~{n}al y Comunicaciones e Ingenier\'{\i}a
Telem\'atica. Universidad de Valladolid. ETS Ingenieros de
Telecomunicaci\'{o}n, Campus Miguel Delibes, E-47011 Valladolid (Spain).}
\author{F. Javier Fraile-Pel\'{a}ez}
\email{fj_fraile@com.uvigo.es}
\affiliation{Dept. Teor\'{\i}a de la Se\~{n}al y Comunicaciones. Universidad de Vigo.\\
ETS Ingenieros de Telecomunicaci\'{o}n, Campus Universitario, E-36310 Vigo (Spain).}
\date{\today}

\begin{abstract}
We analyse the superluminal propagation of narrow-band pulses at resonances in
dissipative media. The output waveform is an attenuated, undistorted,
time-advanced version of the input which can be interpreted as the result of
the interference of two scaled replicas of the input having a positive
relative delay. The proposed analysis applies both to the propagation in a
passive bulk medium at an electronic resonance and in a dielectric waveguide
coupled to a lossy micro-ring resonator. This latter case provides a clear
insight of the underlying physical phenomena.

\end{abstract}

\maketitle

Superluminal propagation is among the most striking phenomena associated to
the propagation of electromagnetic waves. It has been known for long that it
is possible to attain group velocities in excess of $\emph{c}$ when
propagation takes place in anomalous dispersive media \cite{Brillouin}. A
large number of systems have been studied, both theoretically and
experimentally, which hold group velocities larger than $\emph{c}$. Garret and
McCumber \cite{Garrett} considered a smooth pulse propagating through a
dispersive absorbing medium and showed that pulse peak moves at the group
velocity classically defined, even if greater than $c$ or negative. An
experimental confirmation was given in \cite{Chu}. Superluminal group velocity
was studied in transparent media with inverted population by Chiao
\cite{Chiao}. Steinburg \emph{et al.} investigated photon tunneling time
\cite{Steinberg93} and superluminal propagation with negative group velocity
in a medium with a gain doublet \cite{Steinberg}. Bolda \emph{et al.
}considered propagation with negative group velocity due to a nearby gain
line. \cite{Bolda} Later, Dogariu \emph{et al.} demonstrated superluminal
pulse propagation by transparent linear anomalous dispersion created through
the use of two close Raman gain peaks \cite{Dogariu}, and Schweinsberg
\emph{et al.} observed superluminal (as well as slow) pulse propagation in an
erbium-doped fiber (EDF) \cite{Schweinsberg}. More recently, Jiang \emph{et
al.} have given an experimental confirmation that superluminality in an active
Raman medium can be brought about by a single single frequency pump field.
\cite{Jiang}

Although superluminal behavior might seem unphysical, it is always found to be
the result of some sort of artifact, so that the group velocity being greater
than $c$ or negative at certain frequencies does never imply an affront to
relativity. In \cite{Peatross}, a frequency-varying concept of group velocity
is chosen that maintains the meaning of the function $d\omega
/d[\operatorname{Re}(k)]$ even for pulses that experience considerable
distortion through propagation;\ an experimental demonstration using this
description is presented in \cite{Talu}. From a mathematical point of view,
absortion or gain resonances are responsible for the shaping of the linear
dispersive properties of the medium which permit to produce spectral regions
with associated superluminal or negative group delays in bulk dielectric
media.  On the other hand, superluminal propagation in metastructures such as
those based on coupled microring resonator (CMR) has also been predicted
\cite{Boyd}, so the question immediately arises of whether some connection
exists, and to what extent, between both phenomena. As we shall presently see,
the basic physics of the superluminal problem\textbf{ }can be very
insightfully addressed through the analysis of a simple CMR structure having a
lossy ring waveguide. We will show that the model of interference of the input
pulse with scaled self-replicas not only explains the superluminal operation
of the CMR and similar structures, but can also be translated to the context
of propagation through dielectric atomic (bulk) media near the absorption
resonances. Indeed, superluminal linear propagation in such atomic media can
be explained in terms of the interference of the input electromagnetic signal
with the echoes produced by the resonant coupling with the medium via the
linear polarization. We thus focus on the lossy CMR as a specially
illustrative sample system.

When only one mode of the electromagnetic (EM) field is involved, propagation
problems can be addressed using a scalar approach. Two representative examples
are the propagation of plane waves in homogeneous media (the vectorial
formalism in \cite{Peatross}, for example, is in practical terms restricted to
this case) and that of fields in single-mode waveguiding structures. The pulse
propagation problem with a plane wave is entirely equivalent to that of a
waveguide mode, with the only proviso that the modal field profile across the
waveguide is invariant for all frequencies in the pulse spectrum. This
condition virtually applies in all cases of interest, even with ultrashort
pulses containing very few cycles of the carrier. Therefore, in order to
simplify the results, we use a general approach using abstract scalar signals
and linear systems for the discussion of the propagation problems. This does
not restrict the validity of the results and shows that the same principles
are appliable to any linear systems.

We thus consider a generic superluminal system which may consist of either an
atomic medium of finite length or a section of a suitable photonic system. We
assume a time-localized propagating signal with a complex envelope (analytical
signal) $x(t)$. We call $X(\Omega)$ the Fourier transform (FT) of $x(t),$ with
$\Omega=\omega-\omega_{0}$ and $\omega_{0}$ the carrier frequency. We will
denote $y(t)$ the complex envelope of the output signal, $Y(\Omega)$ its FT,
and $H(\Omega)=Y(\Omega)/X(\Omega)$ the system response. Our study deals with
superluminal propagation in resonant structures, so we will assume that
$H(\Omega)$ has a resonance at $\Omega=0$. The \textquotedblleft superluminal
condition\textquotedblright\ is then defined as%

\begin{equation}
\tau_{g}\equiv-\frac{d}{d\Omega}[\arg H(\Omega)]_{\Omega=0}<\tau_{0},
\label{retardos}
\end{equation}
including the case $\tau_{g}<0$. For an atomic medium, $\tau_{0}$ is the group
delay experienced during the propagation in a medium of the same thickness and
the same properties as the one considered, but with the resonance at
$\omega_{0}$ removed from its response. For a CMR such as that shown in Fig.
3, $\tau_{0}$ would correspond to the straight waveguide alone, without the
ring waveguide.

As seen below [c.f. expression (\ref{55})], the superluminal effect in this
case arises from a distortion caused when the propagation in the medium
essentially produces a first replica of the input pulse with a delay $\tau
_{0},$ and a second replica with a delay $\tau_{0}+\Delta\tau.$ It is
necessary that the input pulsewidth be greater than $\Delta\tau$ so that the
second replica can annihilate part of the first without significant
distorison. As a result, the output pulse is apparently time-advanced with
respect to the first replica. Calling $h(t)=$FT$^{-1}[H(\Omega)]$, \emph{i.e.}
the system impulse response, this simply means that%

\begin{equation}
h(t)=a\delta(t-\tau_{0})-b\delta(t-\tau_{0}-\Delta\tau), \label{3}%
\end{equation}
where $a$ and $b$ are real positive constants. Taking the FT of Eq. (\ref{3})
yields $H(\Omega)=a\exp(-i\Omega\tau_{0})-b\exp(-i\Omega\lbrack\tau_{0}%
-\Delta\tau]).$ If the spectrum of $x(t)$ is much narrower than the resonance
peak, the phase is given by%

\begin{equation}
\arg H(\Omega)\simeq-\Omega\tau_{0}+\arctan\frac{b\Omega\Delta\tau}{a-b}.
\label{5}%
\end{equation}

From expression (\ref{5}), the relative group delay relative, defined as the
delay increase wih respect to that of the medium or metastructure without the
resonance, is given by%

\begin{equation}
\tau_{g}-\tau_{0}=-\frac{b\Delta\tau}{a-b}. \label{55}%
\end{equation}

\begin{figure}[ptb]
\includegraphics[width=8.5cm]{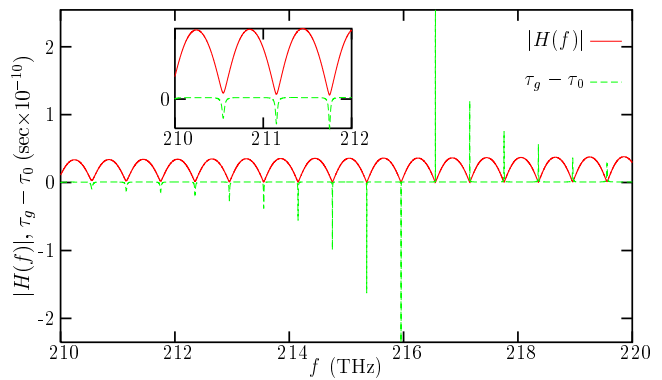}
\caption{Amplitude frequency response and relative group delay for the ring
geometry described in the text.}%
\label{fig1}%
\end{figure}

For superluminal propagation, it is necessary that $b<a.$ Eq. (\ref{55}) also
shows that, as $a\rightarrow b,$ the magnitude of the delay $\tau_{g}$
increases and the amplitude, given by $|H(\Omega)|\simeq a-b,$ decreases.

We analyse a 1-ring \textquotedblleft scissor\textquotedblright\ (side-coupled
integrated spaced-sequence of resonators) as shown in Fig. 3. With an ideal
lossless ring waveguide, it acts as an all-pass optical filter. However, the
presence of radiation losses due to the curvature results in the appearance of
a resonance in the transmission curve. Such losses play the same role as the
absorption losses in an atomic medium, which, in the last analysis, are due to
the electromagnetic energy radiated away by spontaneous emission. The response
of the structure is given by \cite{Boyd}%

\begin{equation}
H(\Omega)=\frac{\theta-\sigma e^{-ikl}}{1-\sigma\theta e^{-ikl}}, \label{H}%
\end{equation}
where $H(\Omega)$ relates the complex envelopes of the output and input modal
electric fields, $0<\theta<1$ is the real transmission coefficient of the
directional coupler (assumed lossless), $l$ is the length of the ring
waveguide, $0<\sigma<1$ is the attenuation factor due to  the radiation losses
in the curved sections, and $k=(\omega/c)\bar{n}$ is the real propagation
constant of the waveguide, with $\bar{n}$ being the modal index and
$\omega=\omega_{0}+\Omega$. The carrier frequency is chosen at a resonance of
the structure: $\omega_{0}l=m2\pi,$ with $m$ an integer, thus $kl=m2\pi
+\Omega\bar{n}l/c.$We then see that, if $\sigma\approx\theta<<1$ (but
$\sigma<\theta$), expression (\ref{H}) simplifies to yield, by inverse
transformation, the result (\ref{3}):%

\begin{equation}
h(t)\simeq\theta\delta(t)-\sigma\delta(t-\Delta\tau),
\end{equation}
with $\Delta\tau=l\bar{n}/c$ Thus, $\theta\equiv a$ and $\sigma\equiv b.$
[Note that, in this case, $\tau_{0}=0$ as the straight, unloaded waveguide is
immaterial in the model of Eq. (\ref{H}).]

\begin{figure}[ptb]
\includegraphics[width=8.5cm]{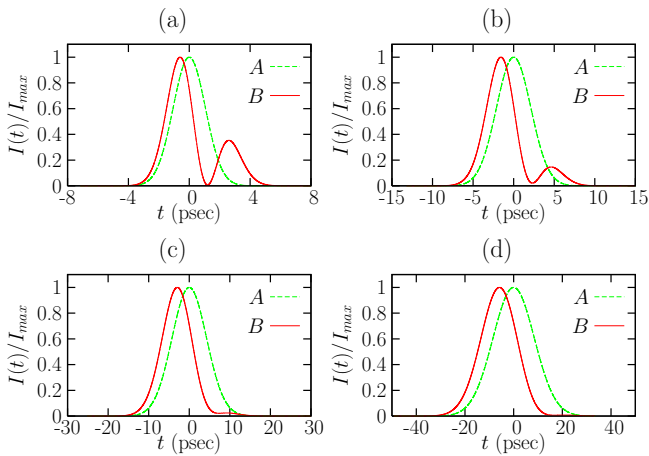}
\caption{Comparison of the normalized reference ($A$) and output ($B$)
waveforms in the ring geometry for (a) $1.5$, (b) $3$, (c) $6$ and (d) $12$
psec $1/e$ input Gaussian pulses. $\Delta\tau\simeq1.6$ psec.}%
\label{fig2}%
\end{figure}

We study the propagation of optical Gaussian pulses using the
Transmission-Line Model method \cite{TLM} to solve the time-domain Maxwell
equations in a two-dimensional geometry. We compare two structures: A
waveguide section and the same waveguide loaded with a micro-ring resonator
built with two straight segments of length $15.25$ $\mu$m joining two
semicircles of $18$ $\mu$m radius. This design permits a better control of the
coupling from the waveguide to the ring. For all the (straight and curved)
waveguide sections, the core refractive index is $n_{1}=3.361$, the cladding
refractive index is $n_{2}=3.168$ and the width of the core is $0.6$ $\mu$m.
The separation of the guides in the coupling region is $0.25$ $\mu$m.

According to our model, the ideal conditions for the observation of
superluminal propagation in the microring resonator structure are defined by a
large coupling to the ring so $\theta$ is small and, simultaneously, large
enough losses so $\sigma$ is comparable to (but smaller than) $\theta$. We
first inject a very short $5$ fsec pulse in order to compute the transfer
function. We compare the signal at a given distance from the ring output with
the reference signal at the same plane propagating in the unloaded waveguide;
these are considered the output and input signals, respectively. Figure
\ref{fig1} shows the amplitude response and the net group delay, computed from
the phase response. The loss mechanism is supplied by the radiation in the
curved sections which increases with frequency. The net group delay is
negative, corresponding to superluminal propagation, up to a frequency limit
when the loss increase sets $\theta>\sigma$. From that point onwards,
subluminal propagation at the resonances is found. As we approach the critical
frequency, the magnitude of the net group delay becomes larger, in agreement
with Eq. \eqref{retardos}, and the resonances become sharper.

\begin{figure}[ptb]
\includegraphics[width=8.5cm]{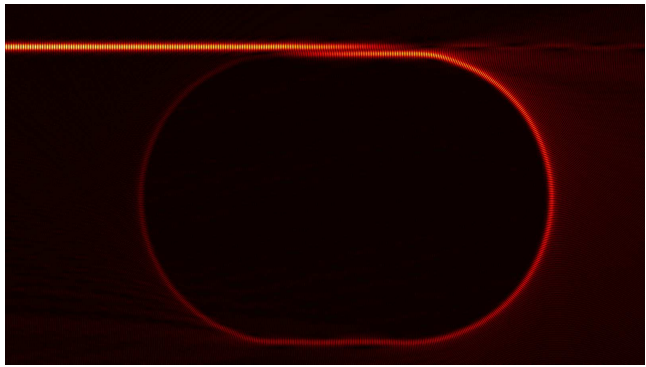}
\caption{Field amplitude distribution $|E(x,z,t=t_{0})|$ at $t_{0}$ when the
leading tail of a $6$ psec Gaussian pulse is entering the ring.}%
\label{fig3}%
\end{figure}

Figure \ref{fig2} compares the normalized reference ($A$) and output ($B$)
signal waveforms for four values of pulse duration: $1.5$, $3$, $6$ and $12$
psec. The carrier frequency of $210.55$ THz is tuned close to the center of
the leftmost resonance in Figure \ref{fig1}. The round-trip time in the ring
is approximately $1.6$ psec. We observe how, as the pulse spectrum narrows in
relation with the resonance bandwidth, the propagation distortion disappears.
The relation between the output/input peak signal levels ranges from $-16$ dB
in case $(a)$ to $-27$ dB in case $(d)$. Figure \ref{fig3} shows the field
amplitude $|E(x,y,t=t_{0})|$ distribution at time $t_{0}$ when the leading
tail of the $6$ psec pulse is entering the ring. For sufficiently long pulses,
the shape of this distribution remains constant until the pulse finally leaves
the structure, given a quasi-stationary picture.

We now turn to the atomic media, where superluminality occurs for narrow-band
pulses near a resonance. The dielectric susceptibiliy can be approximated by%

\begin{equation}
\chi\simeq\chi_{0}-\chi_{r}\frac{i+\Delta}{1+\Delta^{2}},
\end{equation}
where $\chi_{0}$ (real and constant) is the background contribution to $\chi$
from all the resonances above the one considered, and $\Delta\equiv
2\Omega/\gamma$ is the detuning factor normalized to the resonance
width $\gamma/2.$ We will now see that the concept of retarded interference
still applies. The key point is that the role of the ring waveguide in the CMR
structure is played by the electric polarization in the atomic medium. The
atomic polarizability acts as an absorber and retarded reemitter of the
propagating photons. Under certain conditions to be derived next, such
function yields superluminal behavior analogous to that of the electromagnetic
field when fed back into the straight waveguide through the coupler in the CMR.

Starting from the wave equation for the electric field in the slab, $E(z,t),$
denoting $\tilde{E}(z,t)$ the complex field envelope, and using the
slowly-varying envelope approximation (SVEA), one obtains%

\begin{equation}
\frac{\partial\tilde{E}(z,t)}{\partial z}+\left(  \frac{n_{0}}{c}%
-\frac{2\alpha}{\gamma}\right)  \frac{\partial\tilde{E}(z,t)}{\partial
t}=-\alpha\tilde{E}(z,t), \label{eol}%
\end{equation}
where $n_{0}=(1+\chi_{0})^{1/2}$ is the background index of the material,
$\alpha=\chi_{r}\bar{\omega}/(2n_{0}c)$ is the (strong) amplitude attenuation
factor, with $\bar{\omega}$ the carrier frequency, tuned to that of the atomic
resonance. Eq. (\ref{eol}) is derived by considering the frequency dependence
of the resonant polarization up to first order in $\omega-\bar{\omega},$ which
is valid for sufficiently slow (spectrally narrow) pulses. From Eq.
(\ref{eol}) it follows that the resonant group velocity is given by%

\begin{equation}
v_{g}^{-1}=\frac{n_{0}}{c}-\frac{2\alpha}{\gamma},
\end{equation}
where the negative term modifies the background, nondispersive value
$c/n_{0},$ and may lead to negative group velocities. There is a certain
resemblance between this situation and the description of the lossless barrier
tunneling given by Winful \cite{Winful}, where the apparent superluminal
tunneling of narrowband pulses through a barrier is shown to be caused by the
modulation of a standing wave. In our present dissipative, quasistatic
problem, we see that the peak pulse does indeed not propagate superluminality,
but arises from an interference phenomenom.

Now, using the transformations $\tau=t-zn_{0}/c$ and $\zeta=z,$ an equation is
obtained for the amplitude $\tilde{E}^{\prime}(\zeta,\tau)\equiv\tilde
{E}(\zeta,\tau+\zeta n_{0}/c)$ in the frame moving at the background group
velocity. Further, except for the exponential attenuation of the transmited
pulse, the spatial and temporal variations are very slow: The global picture
of the evolution of the optical field in the medium is analogous to that given
in Figure \ref{fig3} for the CMR case. So, by writing $\tilde{E}^{\prime
}(\zeta,\tau)=\exp(-\alpha\zeta)f(\zeta,\tau)$ the following equation is
obtained for the slowly varying amplitude $f(\zeta,\tau)$:%

\begin{equation}
\frac{\partial f}{\partial\zeta}=\frac{2\alpha}{\gamma}\frac{\partial
f}{\partial\tau}. \label{f}%
\end{equation}

In order to bring the result calculations into the formalism of Eqs.
(\ref{3})--(\ref{55}), we approximate Eq. (\ref{f}) by a difference equation,
which can be done because $f(\zeta,\tau)$ is a slow function, with the proviso
that the slab thickness $L$ is sufficiently small. Writing $\partial f(\zeta,\tau
)/\partial\zeta\simeq\lbrack f(L,\tau)-f(0,\tau)]/L$ and $\partial
f(\zeta,\tau)/\partial\tau\simeq\lbrack f(0,\tau)-f(0,\tau-\Delta\tau
)]/\Delta\tau$, where $\Delta\tau=Ln_0/c$, we finally come to the result%

\begin{equation}
\tilde{E}(L,t)\simeq a\tilde{E}(0,t)-b\tilde{E}(0,t-\Delta\tau), \label{ef}%
\end{equation}
with%

\begin{equation}
a=e^{-\alpha L}\left(  1+\frac{2\alpha L}{\gamma\Delta\tau}\right)  ,\quad
b=e^{-\alpha L}\frac{2\alpha L}{\gamma\Delta\tau}.
\end{equation}
Expression (\ref{ef}) is entirely equivalent to expression (\ref{3}), with
$x(t)\equiv\tilde{E}(0,t)$ and $y(t)\equiv\tilde{E}(L,t).$ We see that
$|b|<|a|$ and $|b|\rightarrow|a|$ as $(\gamma\Delta\tau)/(2\alpha L)\rightarrow 0$, which is the same regime as that 
 considered for the CMR case. Figure 4.a shows the input and output normalized
waveforms for a $58$ psec $1/e$ width Gaussian pulse propagating in $9.5$
$\mu$m absorbing slab with the medium parameters obtained from \cite{Chu}. The
output waveforms obtained using the analytical expression
$E(L,t)=E(0,t-L/v_{g})\exp(-\alpha L)$ and the difference equation \eqref{ef}
are displayed in Figure 4.b illustrating excellent agreement.

This work has been supported by the Spanish MEC and FEDER, grant no.
TEC2007-67429-C02-01 and 02.

\begin{figure}[ptb]
\includegraphics[width=8.5cm]{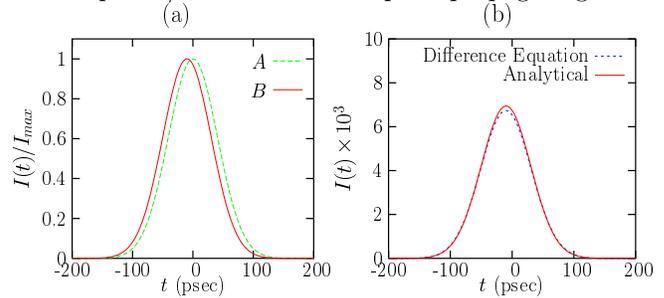}
\caption{(a) Normalized input and output waverforms for the superluminal
propagation of a Gaussian input $58$ psec pulse in a $9.5$ $\mu$m thick medium
with parameters obtained from Ref. \cite{Chu}. (b) Output waveforms as
obtained from the anaylitical solution and the difference equation
\eqref{ef}.}%
\label{fig4}%
\end{figure}

\end{document}